\documentclass[twocolumn]{aastex631}

\newcommand{\G}{\mathcal{G}}

\newcommand{\appropto}{\mathrel{\vcenter{\offinterlineskip\halign{\hfil$##$\cr\propto\cr\noalign{\kern2pt}\sim\cr\noalign{\kern-2pt}}}}}

\newcommand{\msun}{\ensuremath{\,{\rm M_\Sun}}}
\newcommand{\rsun}{\ensuremath{\,{\rm R_\Sun}}}

\newcommand{\degree}{\ensuremath{\,^{\circ}}}

\usepackage{url}
\usepackage{graphicx}
\usepackage{amsmath,amstext}
\usepackage[T1]{fontenc}

\usepackage{amsmath}
\usepackage{float}
\usepackage{longtable}
\usepackage{rotating}
\graphicspath{{./}{Figures/}}

\shorttitle{The Orbital Architecture of Qatar-6: A Fully Aligned 3-Body System?}
\shortauthors{Rice et al.}

\begin{document}
\title{The Orbital Architecture of Qatar-6: A Fully Aligned 3-Body System?}

\author[0000-0002-7670-670X]{Malena Rice} 
\altaffiliation{51 Pegasi b Fellow}
\affiliation{Department of Physics and Kavli Institute for Astrophysics and Space Research, Massachusetts Institute of Technology, Cambridge, MA 02139, USA}
\affiliation{Department of Astronomy, Yale University, New Haven, CT 06511, USA}

\author[0000-0002-7846-6981]{Songhu Wang} 
\affiliation{Department of Astronomy, Indiana University, Bloomington, IN 47405, USA}

\author[0000-0002-4836-1310]{Konstantin Gerbig} 
\affiliation{Department of Astronomy, Yale University, New Haven, CT 06511, USA}

\author[0000-0002-0376-6365]{Xian-Yu Wang} 
\affiliation{National Astronomical Observatories, Chinese Academy of Sciences, Beijing 10010, China}
\affiliation{University of the Chinese Academy of Sciences, Beijing, 100049, China}

\author[0000-0002-8958-0683]{Fei Dai} 
\altaffiliation{NASA Sagan Fellow}
\affiliation{Division of Geological and Planetary Sciences,
1200 E California Blvd, Pasadena, CA, 91125, USA}
\affiliation{Department of Astronomy, California Institute of Technology, Pasadena, CA 91125, USA}

\author[0000-0003-0298-4667]{Dakotah Tyler} 
\affiliation{Astronomy Department, 475 Portola Plaza, University of California, Los Angeles, CA 90095, USA}

\author[0000-0002-0531-1073]{Howard Isaacson} 
\affiliation{Department of Astronomy, University of California Berkeley, Berkeley CA 94720, USA}
\affiliation{Centre for Astrophysics, University of Southern Queensland, Toowoomba, QLD, Australia}

\author[0000-0001-8638-0320]{Andrew W. Howard} 
\affiliation{Department of Astronomy, California Institute of Technology, Pasadena, CA 91125, USA}

\correspondingauthor{Malena Rice}
\email{malena.rice@yale.edu}

\begin{abstract}
The evolutionary history of an extrasolar system is, in part, fossilized through its planets' orbital orientations relative to the host star's spin axis. However, spin-orbit constraints for warm Jupiters -- particularly in binary star systems, which are amenable to a wide range of dynamical processes -- are relatively scarce. We report a measurement of the Rossiter-McLaughlin effect, observed with the Keck/HIRES spectrograph, across the transit of Qatar-6 A b: a warm Jupiter orbiting one star within a binary system. From this measurement, we obtain a sky-projected spin-orbit angle $\lambda=0.1\pm2.6\degree$. Combining this new constraint with the stellar rotational velocity of Qatar-6 A that we measure from TESS photometry, we derive a true obliquity $\psi=21.82^{+8.86}_{-18.36}\degree$ -- consistent with near-exact alignment. We also leverage astrometric data from \textit{Gaia} DR3 to show that the Qatar-6 binary star system is edge-on ($i_{B}=90.17^{+1.07}_{-1.06}\degree$), such that the stellar binary and the transiting exoplanet orbit exhibit line-of-sight orbit-orbit alignment. Ultimately, we demonstrate that all current constraints for the 3-body Qatar-6 system are consistent with both spin-orbit and orbit-orbit alignment. High-precision measurements of the projected stellar spin rate of the host star and the sky-plane geometry of the transit relative to the binary plane are required to conclusively verify the full 3D configuration of the system.
\end{abstract}

\vspace{-10mm}

\section{Introduction} 
\label{section:intro}


In a planetary system, the stellar obliquity is defined as the angle between the stellar spin axis and the net orbital angular momentum vector of the system. While the true stellar obliquity is not currently possible to pinpoint in most exoplanet systems due to incomplete knowledge of where all planets in each system lie, stellar obliquities can be constrained through measurements of individual planets' orbital configurations. These, in turn, offer evidence of the systems' dynamical histories.

As a transiting exoplanet occults its host star, it produces a distortion in the net Doppler shift measured across the integrated light from the star. This distortion is known as the Rossiter-McLaughlin (R-M) effect \citep{rossiter1924detection, mclaughlin1924some}, and it enables a precise measurement of the sky-projected angle $\lambda$ between the stellar spin axis and the transiting exoplanet's orbit normal.

To date, the angle $\lambda$ has been measured for over 170 transiting planets, revealing a diversity of system architectures \citep{albrecht2022stellar}. However, the vast majority of these spin-orbit measurements have been made in hot Jupiter systems due to their relatively deep and frequent transits. By contrast, relatively few $\lambda$ measurements have been made in systems with wider-orbiting warm Jupiters (e.g. \citealt{Wang2021}), which offer important clues to constrain the dominant formation channels for both hot and warm Jupiters \citep{dawson2018origins, jackson2021observable, rice2022tendency}. For the purposes of this work, we define a ``warm Jupiter'' as a short-period ($P<100$ days), Jovian-mass ($0.3M_J<M_b<13M_J$) exoplanet with star-planet separation $a_b/R_*>11$ such that the planet is ``tidally detached'' -- meaning that it undergoes relatively weak tidal interactions with the host star.

We present a measurement of the Rossiter-McLaughlin effect with the Keck/HIRES instrument \citep{vogt1994hires} across a transit of Qatar-6 A b, which is a warm Jupiter residing in a binary star system. This is the fourth result of our Stellar Obliquities in Long-period Exoplanet Systems (SOLES) survey \citep{rice2021soles, wang2022aligned, rice2022tendency} that is systemically extending the census of $\lambda$ measurements to wider-orbiting, tidally detached exoplanets. It is also the ninth measurement of a warm Jupiter spin-orbit angle in a system with one or more known stellar companions \citep[see the TEPcat catalogue;][]{southworth2011homogeneous}.

Qatar-6 A is a young ($1.02\pm0.62$ Gyr), $V=11.5$ early K-type main sequence star that hosts one known sub-Jovian-mass ($M_{b}=0.668\pm0.066M_J$) planet with star-planet separation $a_b/R_*=12.61\pm0.22$ \citep{alsubai2018qatar}. We demonstrate that Qatar-6 A b is likely at or near alignment with the stellar spin axis of its host star, with a projected spin-orbit angle $\lambda=0.1\pm2.6\degree$ and a true spin-orbit angle $\psi=21.82^{+8.86}_{-18.36}\degree$. Considering the larger-scale architecture of the system, we also show that the Qatar-6 AB stellar binary system is edge-on ($i_{B}=90.17^{+1.07}_{-1.06}\degree$) relative to our vantage point from the Earth. This edge-on configuration hints at a potential alignment between the planetary companion's orbit and the orbit of the binary system.

We first describe our observations and data reduction in Section \ref{section:observations}. Then, we characterize the system by extracting stellar parameters in Section \ref{section:stellar_parameters} and modeling both the 2D and 3D stellar obliquity in Section \ref{section:spinorbitmodel}. In Section \ref{section:binary_alignment}, we constrain the Qatar-6 AB binary orbital properties to demonstrate that the system is edge-on, suggesting that the three masses in the system (two stars and one transiting planetary companion) may lie on mutually aligned orbits. We consider relevant timescales for the dynamical evolution of this system in Section \ref{section:dynamical_timescales}, and we discuss the system's potential formation scenarios in Section \ref{section:formation}. Finally, we provide an overview of our findings and their implications in Section \ref{section:conclusions}.

\section{Observations}
\label{section:observations}

We observed the Rossiter-McLaughlin effect across one full transit of Qatar-6 A b from UT 07:25-13:30 on May 30th, 2022 with the Keck/HIRES spectrograph. We obtained thirty-nine 500-second iodine-imprinted radial velocity (RV) exposures during this time span (Table \ref{tab:rv_data_hires}), with a median signal-to-noise ratio (SNR) of 126. Our observing sequence included approximately $150$ and $70$ minutes of pre- and post-transit observations, respectively, to constrain the RV baseline.  All RV observations were taken using the C2 decker ($14\arcsec \times 0.861\arcsec$, $R = 60,000$) and reduced using the California Planet Search pipeline \citep{howard2010california}.

Conditions were stable throughout most of the observing sequence, with seeing ranging from 1.0\arcsec — 1.2\arcsec\, and a short spike in seeing (1.8\arcsec) around UT 08:25-08:45. During the second half of the transit, from UT 11:30-12:25, the presence of clouds substantially reduced the photon count of each spectrum. This is reflected as inflated error bars in the corresponding RV measurements, as shown in Figure \ref{fig:rv_joint_fit}.

To calibrate our RV observations and characterize the system's stellar parameters, we also obtained a 2030-second iodine-free template spectrum of Qatar-6 A with Keck/HIRES during the same night. This single exposure was centered at UT 06:45 and used the spectrograph's B3 decker ($14.0\arcsec \times 0.574\arcsec$, $R = 72,000$). The Qatar-6 A template spectrum was observed in excellent conditions, with 1.1\arcsec\, seeing and SNR$\sim$200.

\begin{deluxetable}{rrrrr}
\tablecaption{Keck/HIRES radial velocities for the Qatar-6 A b planetary system.\label{tab:rv_data_hires}}
\tabletypesize{\scriptsize}
\tablehead{
\colhead{Time (BJD)} & \colhead{RV (m/s)} & \colhead{$\sigma_{\rm RV}$ (m/s)} & \colhead{S-index} & \colhead{$\sigma_S$}}
\tablewidth{300pt}
\startdata
2459729.816537 & 27.56 & 1.74 & 0.555 & 0.001 \\
2459729.822683 & 28.70 & 1.73 & 0.551 & 0.001 \\
2459729.828597 & 20.15 & 2.30 & 0.557 & 0.001 \\
2459729.835425 & 22.04 & 1.83 & 0.571 & 0.001 \\
2459729.841490 & 24.85 & 1.64 & 0.553 & 0.001 \\
2459729.847693 & 25.06 & 1.64 & 0.561 & 0.001 \\
2459729.854035 & 25.88 & 1.58 & 0.553 & 0.001 \\
2459729.866778 & 20.49 & 2.30 & 0.555 & 0.001 \\
2459729.872900 & 13.84 & 1.69 & 0.554 & 0.001 \\
2459729.879208 & 13.98 & 1.64 & 0.541 & 0.001 \\
2459729.885493 & 9.70 & 1.60 & 0.553 & 0.001 \\
2459729.891962 & 12.33 & 1.68 & 0.544 & 0.001 \\
2459729.898073 & 6.03 & 1.57 & 0.554 & 0.001 \\
2459729.904358 & 11.27 & 1.56 & 0.547 & 0.001 \\
2459729.910630 & -1.82 & 1.66 & 0.554 & 0.001 \\
2459729.916833 & -0.66 & 2.04 & 0.570 & 0.001 \\
2459729.923049 & 2.91 & 1.98 & 0.552 & 0.001 \\
2459729.929438 & 4.36 & 2.65 & 0.558 & 0.001 \\
2459729.936196 & -1.59 & 2.06 & 0.541 & 0.001 \\
2459729.942122 & -1.22 & 2.05 & 0.551 & 0.001 \\
2459729.948545 & 12.43 & 1.73 & 0.542 & 0.001 \\
2459729.954760 & 8.68 & 1.77 & 0.543 & 0.001 \\
2459729.961033 & 7.73 & 1.71 & 0.550 & 0.001 \\
2459729.967294 & 6.28 & 1.66 & 0.549 & 0.001 \\
2459729.973555 & -5.04 & 1.74 & 0.553 & 0.001 \\
2459729.979874 & -11.39 & 2.04 & 0.553 & 0.001 \\
2459729.985847 & -17.88 & 2.52 & 0.547 & 0.001 \\
2459729.99234 & -29.59 & 3.47 & 0.615 & 0.001 \\
2459729.998357 & -34.45 & 5.74 & 0.587 & 0.001 \\
2459730.005718 & -18.23 & 5.48 & 0.455 & 0.001 \\
2459730.011725 & -12.29 & 2.58 & 0.528 & 0.001 \\
2459730.017974 & -27.38 & 2.24 & 0.559 & 0.001 \\
2459730.024097 & -18.28 & 1.92 & 0.561 & 0.001 \\
2459730.030254 & -14.63 & 1.88 & 0.551 & 0.001 \\
2459730.036561 & -23.86 & 2.02 & 0.559 & 0.001 \\
2459730.042881 & -17.00 & 1.92 & 0.550 & 0.001 \\
2459730.049154 & -21.06 & 2.00 & 0.560 & 0.001 \\
2459730.055426 & -22.27 & 1.88 & 0.545 & 0.001 \\
2459730.061792 & -27.15 & 2.16 & 0.533 & 0.001
\enddata
\end{deluxetable}

\section{Stellar Parameters} 
\label{section:stellar_parameters}

We first characterized the stellar properties of Qatar-6 A through a spectroscopic analysis of our iodine-free Keck/HIRES template spectrum. To accomplish this, we applied the machine learning model described in \citealt{rice2020stellar}, which is designed to extract precise stellar atmospheric parameters from Keck/HIRES spectra. Our spectroscopic model is trained on 1,202 FGK spectra from the Spectral Properties of Cool Stars (SPOCS) catalogue \citep{valenti2005spectroscopic, brewer2016spectral} and built on the generative machine learning program \textit{The Cannon} \citep{ness2015cannon}. We applied this model to extract four key stellar properties that were directly characterized in the SPOCS catalog: $T_{\rm eff}$, log$g$, $v\sin i_*$, and [Fe/H].

Input spectra to the \citealt{rice2020stellar} model must be continuum-normalized and shifted to the rest frame for direct comparison with the SPOCS training set. We first fit the continuum baseline of Qatar-6 A using the Alpha-shape Fitting to Spectrum (AFS) algorithm described in \citealt{xu2019modeling}, with $\alpha=1/8$ the span of each echelle order. We then divided our initial spectrum by this baseline model to produce a normalized spectrum. Finally, we cross-correlated the normalized spectrum with the solar atlas provided by \citealt{wallace2011optical} to shift the wavelength solution into the rest frame.

We characterized our uncertainties by training and applying our model separately for each of the 16 echelle orders in the Keck/HIRES spectrum, then finding the standard deviation of our results added in quadrature to the training set uncertainties reported in \citealt{brewer2016spectral}. We excluded the fourth echelle order from this analysis due to a poorly fitted baseline subtraction caused by the close proximity of the 5995 \AA\, Na I absorption line to the edge of the echelle order. Our results, provided in the top portion of Table \ref{table:results}, are comprised of the mean and the associated uncertainty derived for each parameter from the 15 remaining model iterations.

We then used these spectroscopically determined stellar parameters, together with parallax constraints from \textit{Gaia} DR3 and archival photometry, as inputs to derive the mass and radius of Qatar-6 A while placing further constraints on $T_{\rm eff}$, $\log g$, and [Fe/H]. We applied the \texttt{isoclassify} Python package \citep{huber2017isoclassify, huber2017asteroseismology, berger2020gaia} to derive posterior distributions for each stellar parameter by fitting a grid of isochrones to our input constraints. Photometry incorporated within our model included magnitudes from Tycho-2 \citep[B and V bands;][]{hog2000tycho}; 2MASS \citep[J, H, and K bands;][]{cutri20032mass}; \textit{Gaia} DR3 \citep[G, Bp, and Rp bands;][]{brown2022gaiadr3}; and the Sloan Digital Sky Survey \citep[u, g, r, i, and z bands;][]{ahn2012ninth}. We set an uncertainty floor of 0.1 mag in each photometric band to facilitate model convergence. We also used an all-sky dust model, implemented through the \texttt{mwdust} Python package \citep{bovy2016galactic}, to fit for extinction. Our results, which are provided in Table \ref{table:results}, are all in agreement with previously published values within $2\sigma$.

\begin{deluxetable*}{lllll}
\tablecaption{Parameters, Priors, and Results for the Qatar-6 A b Planetary System  \label{table:results}}
\tablehead{}
\tablewidth{300pt}
\startdata
& Keck/HIRES & Photometry + Keck/HIRES & Photometry + Keck/HIRES & TRES \\
& This work, spectroscopic fit & This work, isochrone fit & This work, RM fit & Alsubai+ 2018 \\
& \textit{The Cannon} & \texttt{isoclassify} & \texttt{allesfitter} & \texttt{SPC} \\
\hline
\multicolumn{5}{l}{Stellar Parameters:}\\
$M_*$ (\msun) & - & $0.829^{+0.019}_{-0.022}$ & - & $0.822\pm0.021$ \\
$R_*$ (\rsun) & - & $0.785^{+0.033}_{-0.023}$ & - & $0.722\pm0.020$ \\
$T_{\rm eff}$ (K) & $4895\pm 77$ & $5063\pm42$ & - & $5052\pm66$ \\
$\log{g}$ (cm/s$^2$) & $4.41\pm0.20$ & $4.56^{+0.03}_{-0.04}$ & & $4.64\pm0.01$ \\
$v\sin i_{\star}$ (km/s) & 2.4$\pm$1.2 & - & $2.88^{+0.95}_{-0.66}$ & $2.9\pm0.5$ \\
$[\rm{Fe/H}]$ (dex) & $0.06\pm0.05$ & $0.04\pm0.05$ & - & $-0.025\pm0.093$ \\
\\
\hline
Parameter & Description & Priors & Value \\
\hline \\
\multicolumn{5}{l}{Fitted Parameters:}\\
$R_b / R_\star$&     Planet-to-star radius ratio&  $\mathcal U(0.151;0;1)$   &    $0.1516_{-0.0046}^{+0.0055}$ & \\
$(R_\star + R_b) / a_b$&     Sum of radii divided by orbital semimajor axis&   $\mathcal U(0.077;0;1)$   &    $0.0930_{-0.0023}^{+0.0026}$ & \\
$\cos{i_b}$&   Cosine of the orbital inclination&    $\mathcal U(0.0696;0;1)$   &     $0.0707\pm0.0041$ & \\
$T_{0, b}$& Mid-transit epoch (BJD)-2450000   &  $\mathcal U(8611.50257;8610.5025;8612.5025)$   &      $8611.4968\pm0.0082$ & \\
$P_b$& Orbital period (days)      & $\mathcal U(3.506195;0;10)$   &    $3.506200\pm0.000026$ & \\
$K_b$&  RV semi-amplitude (m/s)    &   $\mathcal U(100;0;200)$   &  $110.8_{-5.5}^{+5.8}$ & \\
$\sqrt{e_b} \cos{\omega_b}$&   Eccentricity parameter 1 &    $\mathcal U(0;-1;1)$   &   $0.130_{-0.13}^{+0.094}$ & \\
$\sqrt{e_b} \sin{\omega_b}$&  Eccentricity parameter 2  &   $\mathcal U(0;-1;1)$   &    $0.139_{-0.12}^{+0.087}$ & \\
$\lambda$&  Sky-projected spin–orbit angle ($\degree$)    &  $\mathcal U(0;-180;180)$   &    $0.1\pm2.6$ & \\
$v \sin i_{*}$&  Sky-projected stellar rotational velocity (km/s)   &    $\mathcal U(2.9;0;20)$   &   $2.88_{-0.66}^{+0.95}$ & \\
$q_{1; \mathrm{TESS}}$ & Transformed limb-darkening coefficient 1 &$\mathcal U(0.5;0;1)$   & $0.73_{-0.28}^{+0.20}$ &   \\
$q_{2; \mathrm{TESS}}$ & Transformed limb-darkening coefficient 2&$\mathcal U(0.5;0;1)$   &$0.61_{-0.36}^{+0.28}$ & \\
$q_{1; \mathrm{RM}}$ & Transformed limb-darkening coefficient 1&$\mathcal U(0.5;0;1)$   & $0.36_{-0.26}^{+0.35}$ & \\
$q_{2; \mathrm{RM}}$ & Transformed limb-darkening coefficient 2&$\mathcal U(0.5;0;1)$   & $0.50\pm0.34$ & \\
\\
\multicolumn{5}{l}{Derived Parameters:}\\
$R_{b}$ &      Planetary radius (R$_{J}$)    & ... &  $1.164_{-0.057}^{+0.063}$ & \\
$M_{b}$&      Planetary mass (M$_{J}$)   &   ... & $0.683_{-0.046}^{+0.050}$ & \\
$a_b/R_\star$ & Planetary semi-major axis over host star radius&... &$12.39\pm0.31$ & \\
$b$ &     Impact parameter     & ... &   $0.852_{-0.043}^{+0.029}$ & \\
$T_{\rm 14}$&    Total transit duration (hours)   & ... &   $1.636_{-0.037}^{+0.039}$ & \\
$i_b$ &   Inclination ($\degree$)    &  ... &   $85.95\pm0.24$ & \\          
$e_b$ &     Eccentricity   &   ... &  $0.051_{-0.030}^{+0.032}$ & \\
$\omega_b$&     Argument of periastron ($\degree$) & ...&    $58_{-32}^{+79}$ & \\
$a_b$ &  Semi-major axis (AU)     & ...&   $0.045\pm0.002$ & \\
$u_\mathrm{1; TESS}$ &Limb-darkening parameter 1, TESS  &...& $0.95\pm0.57$ & \\
$u_\mathrm{2; TESS}$ & Limb-darkening parameter 2, TESS &...&$-0.17_{-0.46}^{+0.58}$ & \\
$u_\mathrm{1; RM}$ &Limb-darkening parameter 1, RM &...&$0.50_{-0.36}^{+0.55}$ & \\
$u_\mathrm{2; RM}$ & Limb-darkening parameter 2, RM &... &$0.00\pm0.38$ & \\
\enddata
\end{deluxetable*}

\section{Obliquity Modeling} 
\label{section:spinorbitmodel}

We used the \texttt{allesfitter} Python package \citep{gunther2021allesfitter} to jointly model our new Rossiter-McLaughlin measurements together with other publicly available datasets for Qatar-6 A. Qatar-6 A was observed by the Transiting Exoplanet Survey Satellite \citep[TESS;][]{ricker2015tess} at a 2-minute cadence during Sectors 50 and 51, and both sectors of data were included within our analysis.\footnote{The TESS data used in this paper can be found in MAST: \dataset[10.17909/t9-nmc8-f686]{http://dx.doi.org/10.17909/t9-nmc8-f686}.} Our model also incorporated published radial velocity data from the TRES spectrograph, drawn from \citealt{alsubai2018qatar}. We corrected for systematic additive offsets between RV datasets by fitting and subtracting off a quadratic function between each dataset. The additive offsets account for any correlated noise, instrumental drift, or astrophysical phenomena on timescales longer than $\sim6$ hours.


The free parameters within our model include the companion's orbital period ($P_b$), transit mid-times ($T_{0, b}$), cosine of the planetary orbital inclination ($\cos{i_b}$), planet-to-star radius ratio ($R_{b}/R_{\star}$), sum of radii divided by the orbital semi-major axis ($(R_{\star}+R_{b})/a_b$), RV semi-amplitude ($K_b$), parameterized eccentricity and argument of periastron ($\sqrt{e_b}\,\cos{\,\omega_b}$, $\sqrt{e_b}\,\sin{\,\omega_b}$), sky-projected spin-orbit angle ($\lambda$), sky-projected stellar rotational velocity ($v\sin i_{\star}$), and four limb-darkening coefficients ($q_{1; \mathrm{TESS}}$, $q_{2; \mathrm{TESS}}$, $q_{1; \mathrm{RM}}$, and $q_{2; \mathrm{RM}}$). Each parameter was initialized  with uniform priors within the bounds provided in Table \ref{table:results}.

\begin{figure*}
    \centering
    \includegraphics[width=0.7\textwidth]{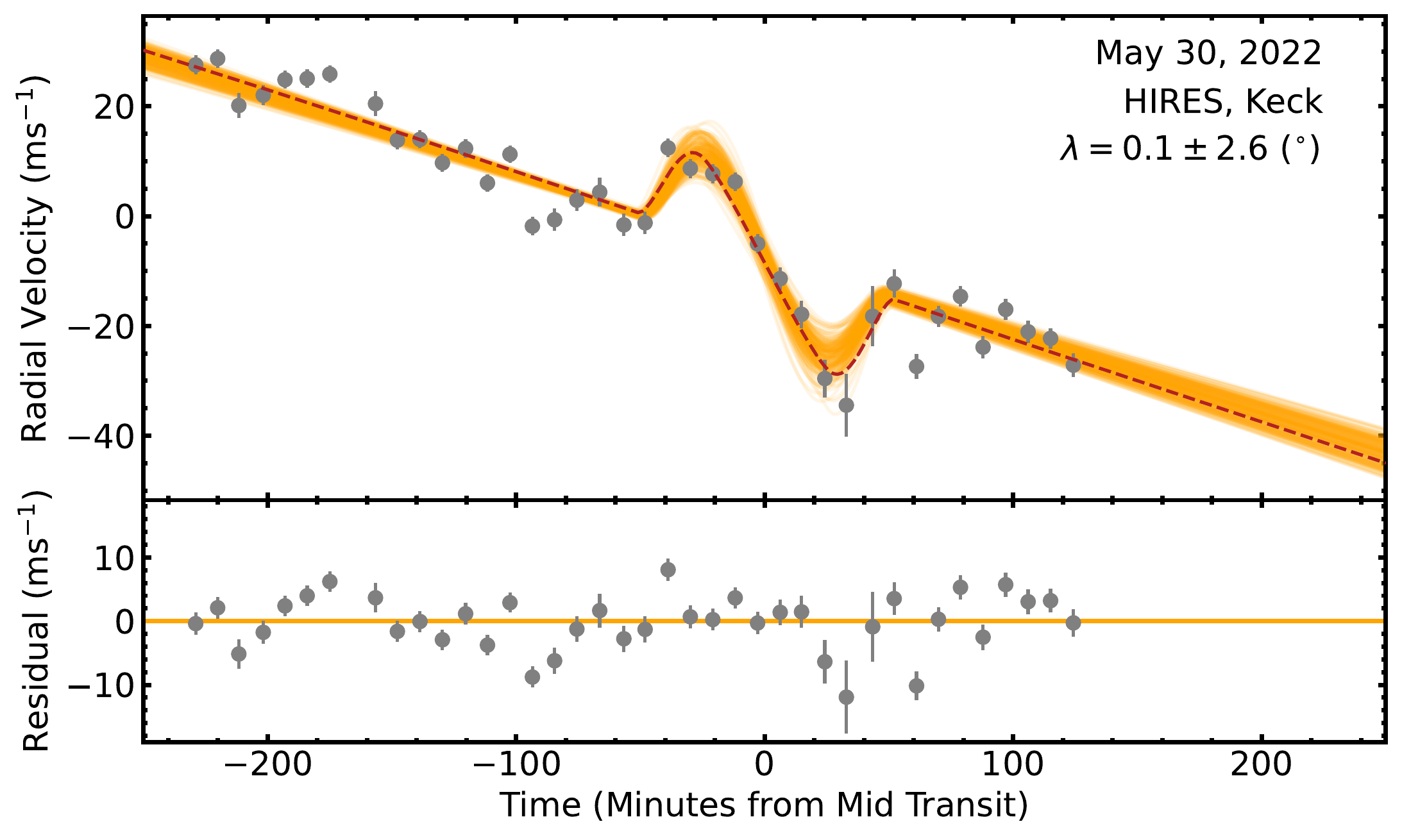}
    \caption{Keck/HIRES observations from the UT 5/30/22 transit of Qatar-6 A b,  with the best-fitting Rossiter-McLaughlin model, corresponding to $\lambda=0.1\pm2.6\degree$, overplotted. The associated residuals from the best-fitting model are provided below.}
    \label{fig:rv_joint_fit}
\end{figure*}

We leveraged an affine-invariant Markov Chain Monte Carlo (MCMC) analysis with 100 walkers to thoroughly sample the posterior distribution for each free parameter, allowing each Markov chain to run over 30$\times$ the autocorrelation length ($\geq500,000$ accepted steps per walker) to ensure convergence. The optimized model results and associated uncertainties are provided in Table \ref{table:results} and displayed in Figure \ref{fig:rv_joint_fit}. We measured a sky-projected stellar obliquity $\lambda=0.1\pm2.6\degree$ for Qatar-6 A b, demonstrating that the system is consistent with alignment. 

Following the routine described in \citealt{Southworth2008}, we adopted the residual-shift method to characterize our uncertainties that may result from unmodeled red noise. We shifted the residual around the best-fit model, point-by-point, until the residuals cycled back to where they originated. After each shift, a new best fit was calculated using the Nelder-Mead algorithm. We ended up with 39 best fits, which is equivalent to the number of RV measurements across our RM curve.  The $\lambda$ value derived from the resulting distribution of fitted values is $\lambda=-0.5\pm2.2^{\circ} $, which is consistent with the result from the \texttt{allesfitter} fit ($\lambda=-0.1\pm2.6^{\circ} $). To be conservative, the latter was used since its error is larger than that of the former.

We also measured the stellar rotation period of Qatar-6 A to determine the 3D stellar obliquity, $\psi$. We employed a Generalised Lomb-Scargle periodogram \citep[GLS;][]{Zechmeister2009} to analyze the TESS light curves from Sectors 50 and 51 and to extract key periodicities.

There are currently two types of light curves provided by the TESS SPOC pipeline: the Simple Aperture Photometry (SAP) light curves and the Pre-search Data Conditioning SAP \cite[PDCSAP;][]{Jenkins2016} light curves. SAP light curves were used in our work, since the stellar rotation signals could be recognized as spacecraft-related systematics that would be removed in the PDCSAP detrending process. The SAP light curves were downloaded using the \texttt{lightkurve} Python package \citep{cardoso2018lightkurve}, and all transits and flagged measurements were masked. 

\begin{figure*}
    \centering
    \includegraphics[width=0.7\textwidth]{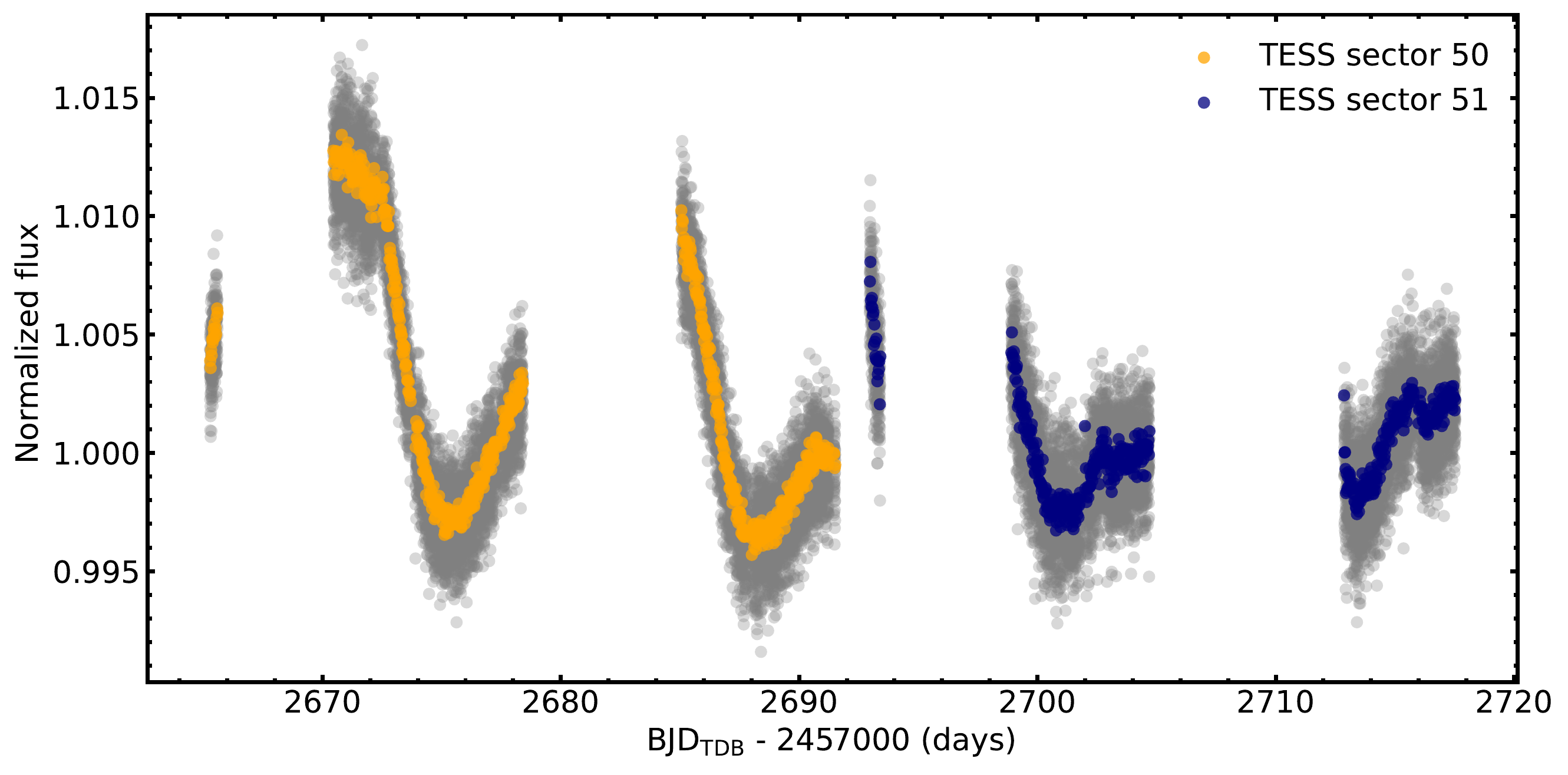}
    \caption{Simple Aperture Photometry (SAP) light curve for Qatar-6 A with transits masked. The star shows significant rotational modulations with a periodicity of $P_* = 12.962\pm0.015$ days based on a Generalised Lomb-Scargle analysis.}
    \label{fig:Rotation}
\end{figure*}

From the reduced SAP light curves, Qatar-6 A shows apparent rotational modulation with period of $P_* = 12.962\pm0.015$ days (see Figure~\ref{fig:Rotation}). Our result is in excellent agreement with the rotation period $12.75\pm1.75$ days derived in \citealt{alsubai2018qatar}. 

We then used this result to derive the stellar inclination $i_*$, following the methods described in \citealt{Masuda2020}. We adopted the Affine-Invariant Monte Carlo Markov Chain method implemented in the \texttt{emcee} Python package \citep{foremanmackey2013} to derive posterior distributions for three independent variables -- $R_*$, $P_*$, and $\cos{i_*}$ -- with measurement-informed priors on $R_*$, $P_*$, and $(2 \pi R_*/P_*)\sqrt{1-\cos^{2}{i_*}}$. From this, we obtained the stellar inclination $i_*=66.8^{+9.7}_{-23.3}\degree$.


Combining the derived stellar inclination $i_*$ with our newly constrained sky-projected spin-orbit angle $\lambda$ and the orbital inclination measurement ($i_b$) derived from our global fit to the Qatar-6 A b system, we calculated the true spin-orbit angle $\psi$ using the relation

\begin{equation}
    \cos \psi = \sin i_* \cos \lambda \sin i_b + \cos i_* \cos i_b.
\end{equation}
We obtained $\psi=21.82^{+8.86}_{-18.36}\degree$, indicating that Qatar-6 A b is consistent with spin-orbit alignment. The large error bars in $\psi$ are primarily driven by the relatively large uncertainty in our measured $v\sin i_*$. This value may be better constrained with higher-resolution spectrographs mounted on large-aperture telescopes amenable to observations of relatively dim stars, such as the Keck Planet Finder \citep{gibson2016kpf}.

\section{Binary System Alignment}
\label{section:binary_alignment} 

The Qatar-6 system contains two stars, Qatar-6 A and Qatar-6 B, together with the companion planet Qatar-6 A b. The relative orbital configurations of the full 3-body system can, therefore, offer clues to the system's most likely formation mechanism. In this section, we leverage astrometric data from \textit{Gaia} DR3 \citep{brown2022gaiadr3} to constrain the orbital properties of the Qatar-6 AB binary star system.

\subsection{Confirming the System's Stellar Multiplicity}
\label{subsection:multiplicity} 

The system's secondary star, Qatar-6 B, was first identified by \citealt{mugrauer2019search}. \citealt{mugrauer2019search} used the \textit{Gaia} DR2 astrometric dataset \citep{Gaia2016, brown2018gaia} to demonstrate that Qatar-6 B is bound to the primary. In this section, we use the updated \textit{Gaia} DR3 dataset to confirm that Qatar-6 B is the only candidate stellar companion to Qatar-6 A. 

We queried for all sources within $10\arcmin$ of Qatar-6 A, then followed the methods of \citealt{el2021million} to vet potential companions. Our search included any sources with parallaxes $\varpi>1$ mas, fractional parallax uncertainties $\sigma_{\varpi}/\varpi<0.2$, and absolute parallax uncertainties $\sigma_{\varpi}<2$ mas. For each source that passed this initial cut, we checked the following three requirements:

\vspace{2mm}
\noindent(i) The sky-projected separation $s$ between the two stars must be less than 1 pc; that is,

\begin{equation}
    \Big(\frac{\theta_s}{\mathrm{arcsec}}\Big) < 206.265 \Big(\frac{\varpi}{\mathrm{mas}}\Big)
\end{equation}
for projected angular separation $\theta_s$ calculated as 

 \begin{equation}
   \cos{(\theta_s)} = (\sin\delta_{p}\sin\delta_{cc}+\cos\delta_{p}\cos\delta_{cc}\cos({\alpha_p - \alpha_{cc}))}.
\end{equation}
The subscript $p$ refers to the primary (Qatar-6 A), whereas the subscript $cc$ refers to a companion candidate. RA and Dec are given as $\alpha$ and $\delta$, respectively.

\vspace{2mm}
\noindent(ii) The parallax of the primary and the candidate companion must be consistent within $b\sigma$, such that

\begin{equation}
   |\varpi_{cc} - \varpi_p| < b \sqrt{\sigma_{\varpi_{cc}}^2 + \sigma_{\varpi_{p}}^2}.
\end{equation}
Following \citealt{el2021million}, we set $b = 3$ for pairs with angular separation $\theta_s>4.0\arcsec$, or $b = 6$ for pairs with $\theta_s<4.0\arcsec$. The weaker threshold for pairs at small angular separation is set to counteract the systematic underestimate of parallax uncertainties at close angular separations \citep{el2021million}.

\vspace{2mm}
\noindent(iii) The two stars must have relative proper motion measurements consistent with a bound Keplerian orbit

\begin{equation}
    \Delta\mu < \Delta\mu_{\rm orbit} + 2\sigma_{\Delta\mu}.
\end{equation}
Here, $\Delta\mu_{\rm orbit}$ is given by 

\begin{equation}
    \Delta\mu_{\rm orbit} =( 0.44\, \mathrm{mas/yr})\,\Big(\frac{\varpi}{\mathrm{mas}}\Big)^{3/2}\Big(\frac{\theta_s}{\mathrm{arcsec}}\Big)^{1/2}.
\end{equation}
This relation, which is drawn from \citealt{el2018imprints}, provides the maximum difference in proper motion for a circular binary orbit with total system mass 5$M_{\odot}$. We use this as a conservative estimate to encapsulate a range of potential candidate stellar companion velocities.

The uncertainty $\sigma_{\Delta\mu}$ in the proper motion difference between the two stellar components is given as

\begin{equation}
    \sigma_{\Delta\mu} = \frac{1}{\Delta\mu}\sqrt{(\sigma_{\mu^*_{\alpha, 1}}^2 + \sigma_{\mu^*_{\alpha, 1}}^2)\Delta{\mu^*_{\alpha}}^2 + (\sigma_{\mu_{\delta, 1}}^2 + \sigma_{\mu_{\delta, 2}}^2)\Delta\mu_{\delta}^2},
\end{equation}
while the proper motion difference $\Delta\mu$ is

\begin{equation}
    \Delta\mu = \sqrt{\Delta{\mu^*_{\alpha}}^2 + \Delta\mu_{\delta}^2}.
\end{equation}
The proper motion differences $\Delta\mu^*_{\alpha}$ and $\Delta\mu_{\delta}$ in the RA and Dec directions, respectively, are calculated as

\begin{equation}
	\Delta{\mu^*_{\alpha}}^2 = (\mu^*_{\alpha,1} - \mu^*_{\alpha, 2})^2
\end{equation}
and

\begin{equation}
	\Delta{\mu_{\delta}}^2 = (\mu_{\delta,1} - \mu_{\delta, 2})^2.
\end{equation}

We note that proper motions reported by \textit{Gaia} DR3 in the RA direction already account for a $\cos{\delta}$ corrective factor,\footnote{See the documentation for \textit{Gaia} source parameters; \url{https://gea.esac.esa.int/archive/documentation/GEDR3/Gaia_archive/chap_datamodel/sec_dm_main_tables/ssec_dm_gaia_source.html}} such that $\mu^*_{\alpha} \equiv \mu_{\alpha}\cos\delta$. Our provided relations have implicitly included this correction within them. For clarity, we include a star superscript on each variable that includes this corrective factor.

A single source passed all three of the tests outlined above. We first compared the projected separation and orientation of the system to confirm that the recovered source is the previously identified companion Qatar-6 B. Then, we verified that the same binary companion was also identified in \citealt{el2021million}, which predicts a low fractional chance alignment probability $R=1.92125073\times10^{-6}$ for the binary pair. We conclude that the previously identified companion Qatar-6 B -- the only candidate companion that we identify for Qatar-6 A -- is likely not a chance alignment.

We also confirmed that the projected separation of the binary is $s<<30,000$ au, a threshold above which \citealt{el2021million} finds that chance alignments dominate over true bound companions. Using \textit{Gaia} DR3, we measured a sky-projected separation $s=482$ au between Qatar-6 A and Qatar-6 B. This separation is similar to, but slightly smaller than, the $s=486$ au separation determined by \citealt{mugrauer2019search} using \textit{Gaia} DR2.

Lastly, we checked each star's Renormalized Unit Weight Error (RUWE) parameter -- a $\chi^2$-based metric provided by \textit{Gaia} to quantify the robustness of the astrometric fit for a star. RUWE $\sim1.0$ typically corresponds to a high-quality single star fit, while RUWE $>1.4$ generally indicates a poor astrometric fit that may result from the presence of an unresolved companion. For Qatar-6 A and B, respectively, \textit{Gaia} DR3 reports RUWE $=1.22$ and RUWE $=1.20$. Although both Qatar-6 stars fall comfortably below the commonly-adopted limit RUWE $<1.4$, the astrometric fit for each star deviates substantially from RUWE $=1.0$. With this caveat in mind, we proceed to further characterize the properties of the binary star orbits.

\subsection{Constraining the Binary Star Orbit}
\label{subsection:binary_alignment} 

\begin{figure}
    \centering
    \includegraphics[width=0.48\textwidth]{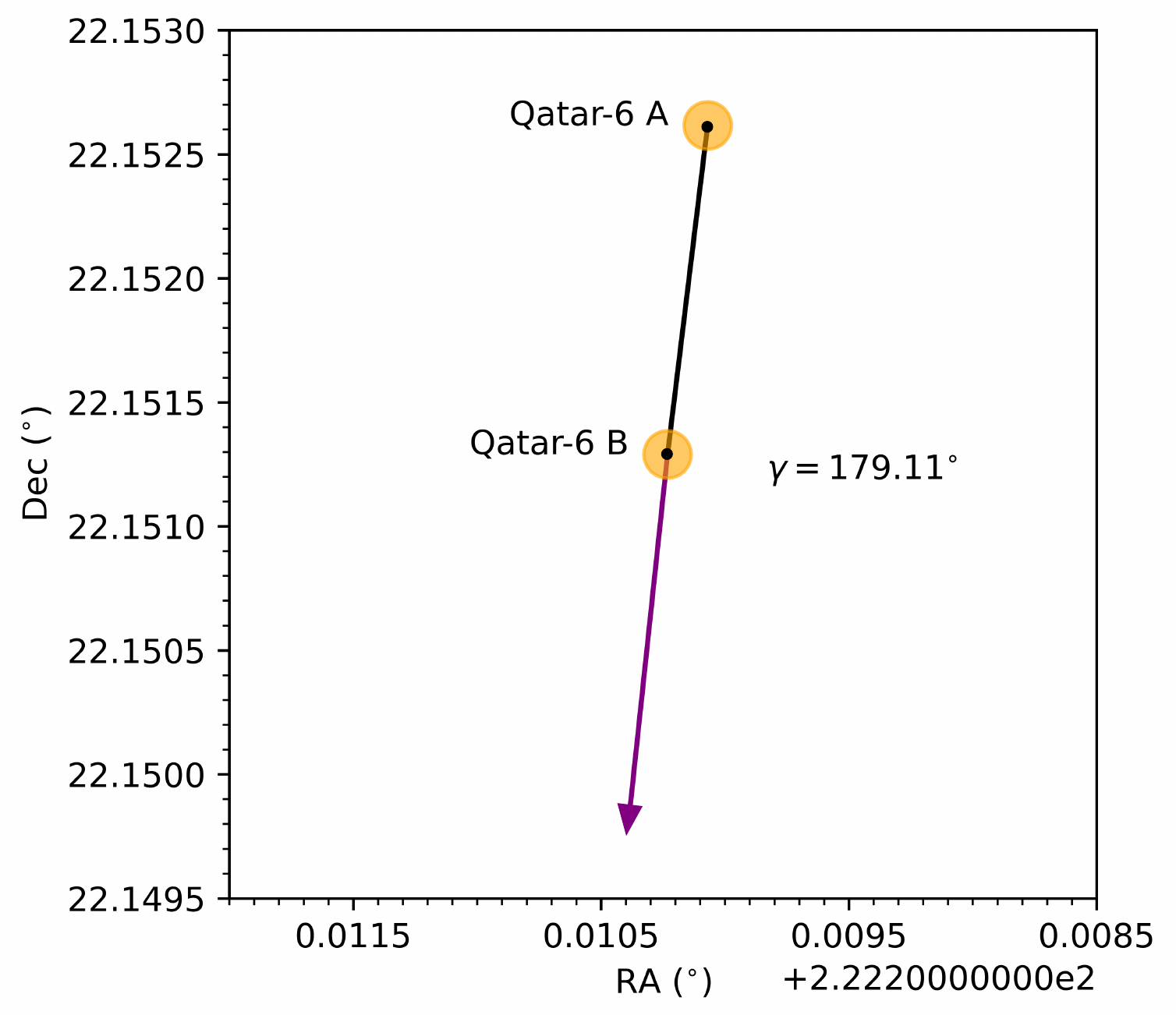}
    \caption{Geometry of the Qatar-6 AB stellar system. We orient our schematic with north upwards and east to the left. The position vector between the two stars (black) and the relative proper motion vector (purple), scaled to demonstrate the geometry of the system, are nearly linear within the sky plane. Qatar-6 B is moving in a sky-projected direction $\gamma=179.11\degree$ away from the primary, Qatar-6 A.}
    \label{fig:gamma_Q6}
\end{figure}

To examine the relative orbital configuration of the Qatar-6 AB binary star system, we first measured the angle $\gamma$ between the position vector connecting the two binary star components and the relative proper motion vector \citep{tokovinin1998distribution, tokovinin2015eccentricity}. We selected the convention that $\gamma=180\degree$ where the secondary star is moving directly in the opposite direction of the primary, whereas $\gamma=0\degree$ where the secondary star is moving directly toward the primary. We obtained $\gamma=179.11\degree$ for Qatar-6 AB, and the geometry of the system is visualized in Figure \ref{fig:gamma_Q6} for reference. This nearly linear configuration indicates that the position and velocity vectors are well-aligned within the sky plane, suggesting an edge-on orbit for the binary system. 

Next, we further constrained the highest-likelihood binary orbits for the Qatar-6 AB system using the \texttt{lofti\_gaia} Python package \citep{pearce2020orbital}. The LOFTI (Linear OFTI) algorithm implemented by \texttt{lofti\_gaia} was designed to constrain the orbital properties of binary star systems based on the linear sky-plane velocity vector of each star measured by the \textit{Gaia} mission. LOFTI builds upon the Orbits For The Impatient (OFTI) Bayesian rejection sampling method \citep{blunt2017orbits} for orbit fitting to short orbital arcs.

We ran LOFTI up to 100,000 accepted orbits using astrometric constraints provided by \textit{Gaia} DR3. Stellar masses $M_{A}=0.829^{+0.019}_{-0.022} M_{\Sun}$ (this work) and $M_{B}=0.244^{+0.013}_{-0.020}M_{\Sun}$ \citep{mugrauer2019search} were adopted for Qatar-6 A and Qatar-6 B, respectively. Because LOFTI requires a symmetric mass uncertainty, we used the larger uncertainties $\sigma_{M_{A}}= 0.022M_{\Sun}$ for Qatar-6 A and $\sigma_{M_{B}}=0.020M_{\Sun}$ for Qatar-6 B.

\begin{figure}
    \centering
    \includegraphics[width=0.48\textwidth]{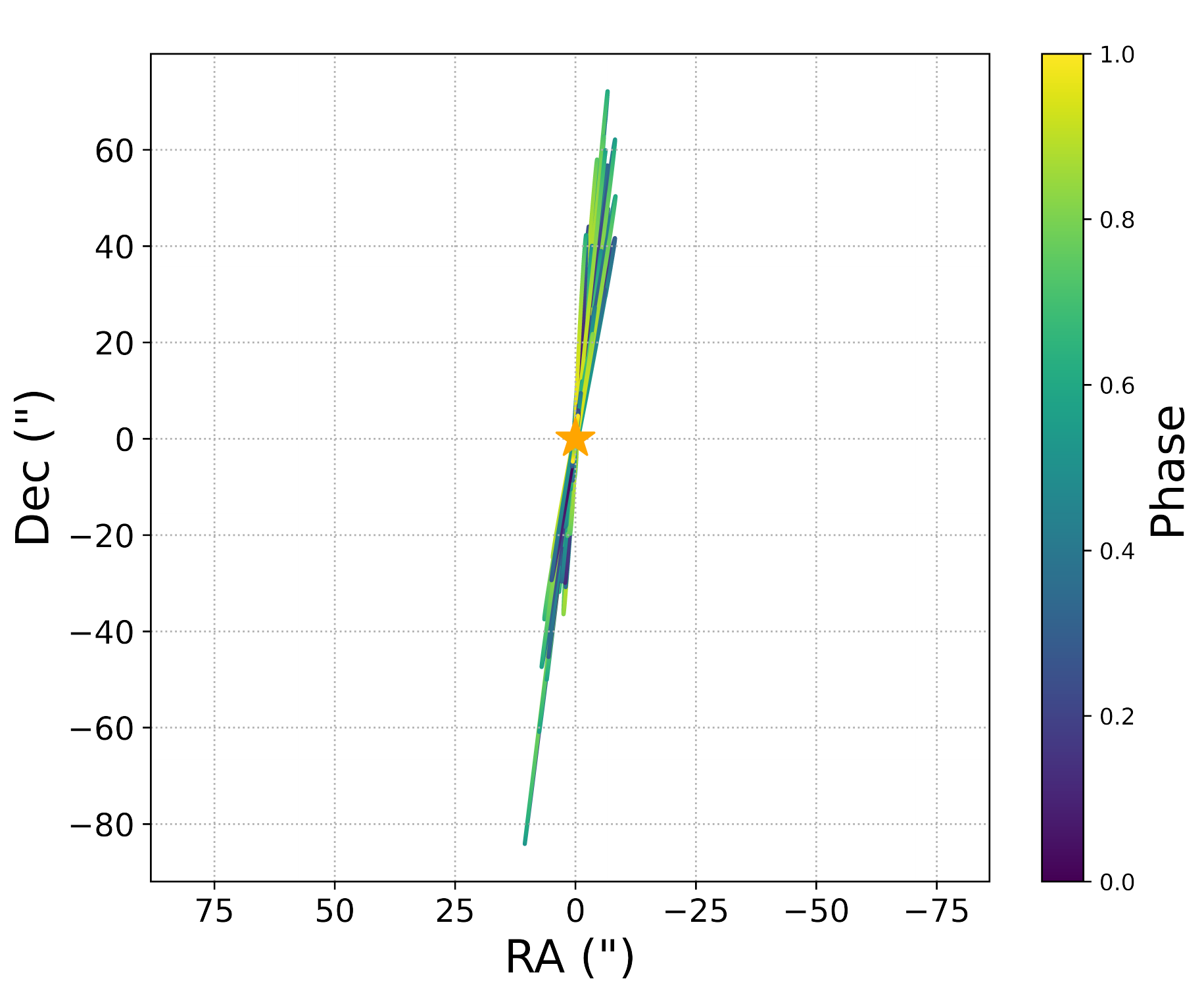}
    \caption{Selection of 1,000 accepted orbits from the posterior of LOFTI orbit fits to the Qatar-6 AB stellar system. All accepted orbits have nearly edge-on inclinations, with $i_{B}=90.17^{+1.07}_{-1.06}\degree$. Here, the RA and Dec of the primary star have been centered at (0,0).}
    \label{fig:sample_orbits}
\end{figure}
\begin{figure*}
    \centering
    \includegraphics[width=0.98\textwidth]{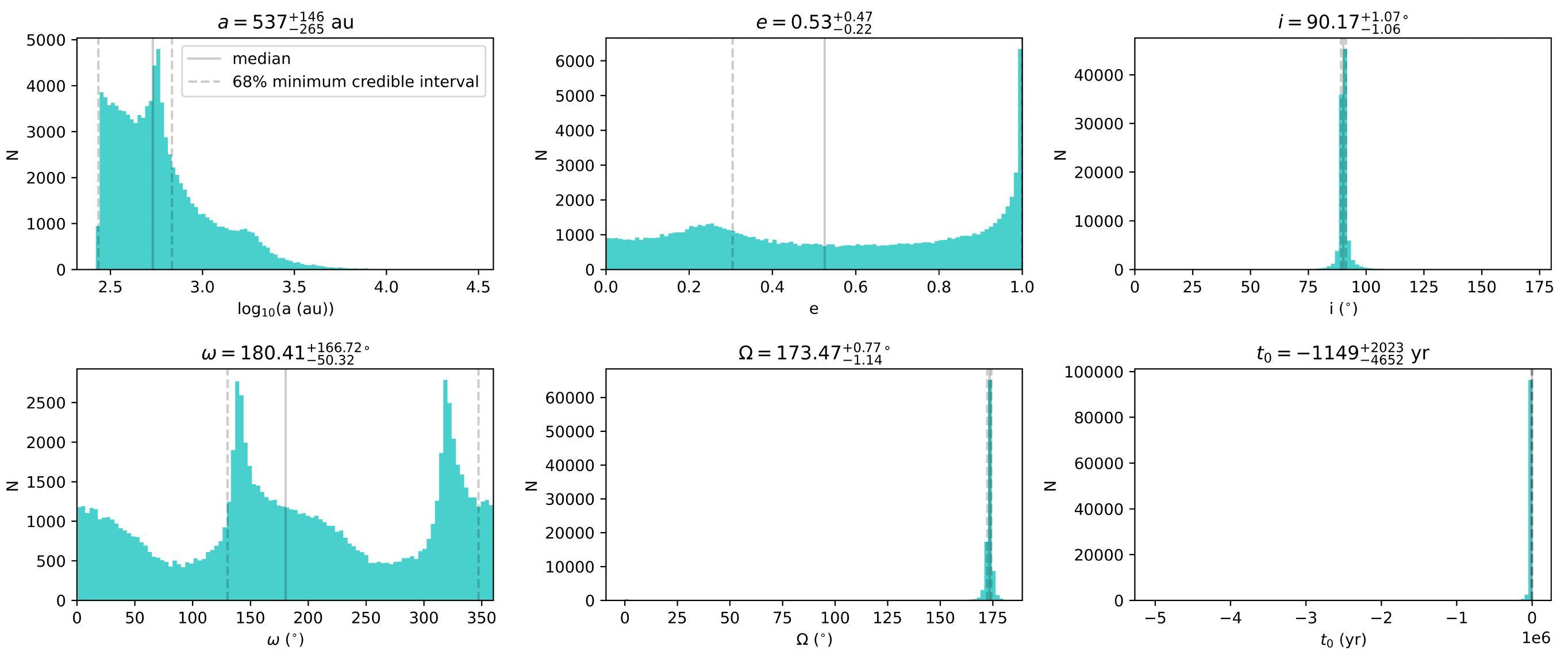}
    \caption{Posteriors from LOFTI fits to the Qatar-6 AB binary star system. The inclination ($i$) and longitude of ascending node ($\Omega$) for the binary system are well-constrained by \textit{Gaia} DR3 astrometry. The median and 68\% minimum credible interval for each parameter are each shown in gray and provided along the top of each panel.}
    \label{fig:histograms}
\end{figure*}

A subsample of 1,000 accepted orbits is shown in Figure \ref{fig:sample_orbits}, demonstrating a clear tendency toward aligned orbits in agreement with our $\gamma$ analysis. The posteriors of our orbit fit are provided in Figure \ref{fig:histograms}. We find that the current set of \textit{Gaia} DR3 astrometry is unable to provide a strong constraint on the system's eccentricity. There is also a degeneracy between position angles $PA=180\degree$ and $PA=360\degree$ due to the relatively small, $2.2\sigma$ difference in parallaxes between the two components, which produces a corresponding degeneracy in the argument of periapsis $\omega$. Regardless, we find a strong preference for an edge-on binary orbit, with inclination $i_{B}=90.17^{+1.07}_{-1.06}\degree$.

Ignoring selection effects \citep[see e.g.,][]{el2018imprints, ferrer2021biases}, the probability distribution function for inclinations $i_B$ drawn from an isotropically distributed set of orbits follows a uniform distribution in $\sin(i_B)$, where $i_B=90\degree$ is defined as an edge-on orbit. In the case that stellar binary orbits are randomly oriented, there would be a $\sim 2 \%$ occurrence rate for chance alignments in $i_B$ within $1.2\degree$. Correspondingly, there is a $\sim 2 \%$ chance that the observed alignment results from a chance alignment among a randomly distributed set of orbits.

\begin{figure*}
    \centering
    \includegraphics[width=0.98\textwidth]{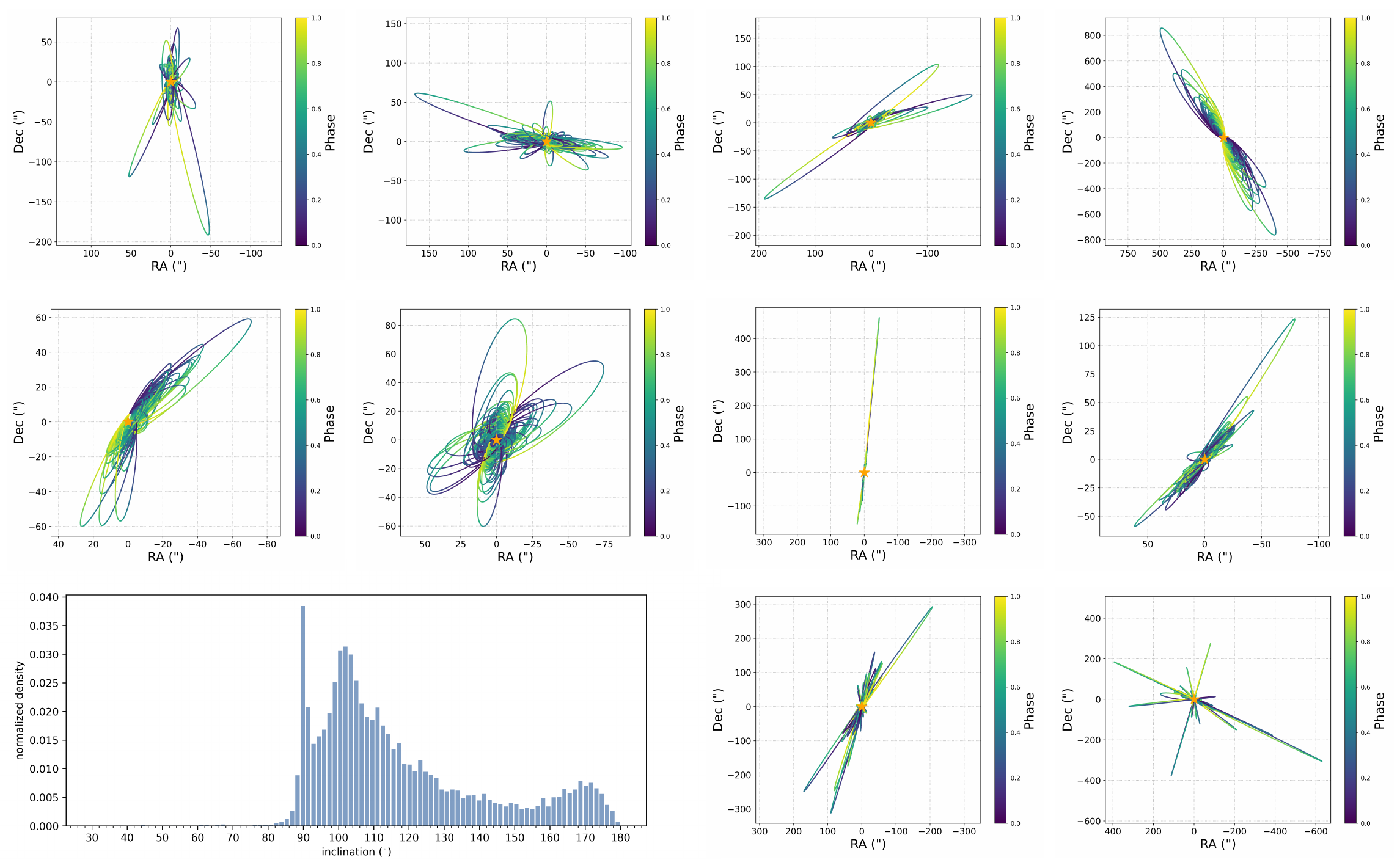}
    \caption{Gallery of 10 comparison systems with similar properties to Qatar-6, together with the normalized density of accepted orbital inclinations across all samples (bottom left). 1,000 accepted orbits, fit to \textit{Gaia} DR3 astrometric constraints, are shown for each system. While we do find a tendency toward edge-on systems, the distribution of accepted orbits in our comparison sample is much broader than that of the Qatar-6 AB binary system.}
    \label{fig:comparison_sample}
\end{figure*}

Selection biases favor relatively edge-on orbits, such that even an isotropically distributed set of orbits (uniform in $\sin(i_B)$) should include an overdensity toward $i_B\sim90\degree$. Furthermore, orbit fitting with little to no orbital coverage suffers from known degeneracies between inclination and eccentricity \citep{ferrer2021biases}. To examine the robustness of our edge-on orbit fit, we produced a comparison sample of ten systems with similar binary separation, parallax, and magnitude properties to that of Qatar-6 AB. This sample includes the ten systems within the \citealt{el2021million} catalogue with the most similar properties to that of Qatar-6 AB based on the metric adopted in \citealt{christian2022possible}. Masses were extracted using the \texttt{isoclassify} Python package in the same configuration described in Section \ref{section:stellar_parameters}, but with an uncertainty floor of 0.5 mag to facilitate convergence. We integrated each comparison system to 1,000 accepted orbits using LOFTI, with results that are shown in Figure \ref{fig:comparison_sample}. This exercise reaffirms the edge-on nature of Qatar-6 AB, which has an inclination distribution that is restricted to a much narrower range of edge-on values than the comparison sample.

Recent population studies have demonstrated that hosts of transiting planets tend to have stellar binary orbits closer to an edge-on configuration than field binary star systems with no known planets, suggesting a systematic trend toward alignment between the binary plane and the planetary orbits \citep{christian2022possible, dupuy2022orbital}. Our results are consistent with this finding: that is, we show that the Qatar-6 AB binary system, which includes one confirmed transiting planet, lies in a precisely edge-on configuration. Combined with previous findings, this suggests that the planet's orbital plane may be closely aligned with the stellar binary orbital plane (a configuration that we refer to as ``orbit-orbit alignment'').

However, we emphasize that the only well-constrained angle in our binary system is inclination. That is, the sky-plane direction in which the planet is transiting is not constrained relative to the plane of the stellar orbit. While the line-of-sight orientation of the system is consistent with alignment, it is not currently possible to measure the sky-plane angle between the planetary orbit and the stellar orbit in most planetary systems. Additional population-wide studies examining the prevalence of orbit-orbit alignment in hot and warm Jupiter systems may further inform the role of stellar multiplicity in the evolution of planetary systems.

\section{Dynamical Timescales}
\label{section:dynamical_timescales}

The timescales of relevant dynamical mechanisms can be compared with the system age to better constrain the past evolution of a given planetary system. In this section, we examine several important timescales at play in the Qatar-6 system: the tidal alignment timescale (Section \ref{subsection:tidal_align_timescale}), the tidal circularization timescale (Section \ref{subsection:tidal_circ_timescale}), the Kozai-Lidov timescale (Section \ref{subsection:kozai_timescale}), the apsidal precession timescale from general relativity (Section \ref{subsection:apsidal_timescale}), and the timescales for changes to each orbital element under the influence of the Qatar-6 B secondary star (Section \ref{subsection:orb_elem_timescale}). We then discuss the joint implications of these timescales in Section \ref{subsection:timescale_implications}.

\subsection{Tidal Alignment Timescale}
\label{subsection:tidal_align_timescale}

\subsubsection{Angular Momentum of the System}

To evaluate the feasibility of tidal alignment within the system, we first compared the orbital angular momentum of Qatar-6 A b with that of the convective layer of Qatar-6 A. In the case that tidal alignment occurs prior to the completion of the orbital decay process (and the subsequent disruption of the Jovian planet), the companion planet's orbit should host more angular momentum than the host star's convective layer. We calculated the angular momentum $L_{CZ}$ of the star's convective zone using the relation

\begin{equation}
    L_{CZ} = \omega\int l^2 dm
    \label{eq:L_cz}
\end{equation}
for distance $l$ from each mass element $dm$ to the spin axis, with $l=r\sin\phi$ and

\begin{equation}
    dm = \rho dV = \rho r^2 \sin\phi dr d\phi d\theta.
\end{equation}
Here, $\omega=v/R_*$ is the angular velocity of the convective layer's rotation, where we assume no shear between layers. We integrated the density $\rho$ over each volume element $dV$ of the convective layer, using spherical coordinates to integrate from the radius of the convective zone boundary ($r=R_{CZ}$) to the full radius of the star ($r=R_*$). The radius of the convective zone boundary was set as $R_{CZ}/R_*=0.69$ based on the models of \citealt{van2012sensitivity} for the mass and age of Qatar-6 A. The density of the convective layer was approximated as uniform, with a total mass $M_{CZ}=10^{-1.35}M_{\odot}$ drawn from the stellar interior models of \citealt{pinsonneault2001mass}. 

Comparing this with the planet's orbital angular momentum, which is given by 

\begin{equation}
    L_{p, orb}=M_b v_b r_b
\end{equation}
at a given orbital distance $r_b$ and momentary velocity $v_b$, we find that $L_{p, orb}/L_{CZ}\sim15$. This excess of angular momentum in the planet's orbit indicates that tidal alignment could feasibly occur within this system.

\subsubsection{Equilibrium Tides}
Qatar-6 A is a cool star ($T_{\rm eff}\sim5063\pm42$ K) that lies well below the Kraft break — a rotational discontinuity at roughly $T_{\rm eff}\sim6100$ K, below which stars typically have convective envelopes \citep{kraft1967studies}. The tidal alignment timescale $\tau_{CE}$ for stars with convective envelopes can be approximated as

\begin{equation}
    \tau_{CE} = \frac{10^{10} \rm{yr}}{(M_b/M_*)^2}\Big(\frac{a_b/R_*}{40}\Big)^{6},
\label{eq:tau}
\end{equation}
where $M_b/M_*$ is the planet-to-star mass ratio \citep{zahn1977tidal, albrecht2012obliquities}. This scaling is calibrated based on the observed synchronization of stellar binaries under the framework of equilibrium tides. Consequently, it includes an implicit assumption that the tidal realignment timescale for planetary systems scales similarly to that of stellar systems.

Based on Equation \ref{eq:tau}, Qatar-6 A b has $\tau_{CE}\sim 1.6\times10^{13}$ years, which is longer than the age of the Universe. Therefore, under the assumption that equilibrium tides well approximate the dynamical behavior of this system, the companion planet's spin-orbit angle likely has not changed significantly since its initial formation. 

\subsubsection{Dynamical Tides}
\label{subsubsection:dynamical_tides}

Alternatively, the system may have been aligned through the dissipation of inertial waves, which are driven by the Coriolis force in a rotating star. The tidal disturbances produced through this mechanism are collectively known as ``dynamical tides''. We focus on a specific mode of the dynamical tide -- known as the ``obliquity tide'', with $m=\pm1$ and $m'=0$ -- that damps only the stellar obliquity without altering the orbital semimajor axis of the companion planet \citep{lai2012tidal}. In this case, the obliquity of a planetary system evolves as

\begin{equation}
\begin{split}
    \frac{d\psi}{dt}\bigg\rvert_{10} = -\frac{3}{4}\frac{k_{2}}{Q_{10}}\Big(\frac{M_b}{M_*}\Big)\Big(\frac{R_*}{a_b}\Big)^5\Omega_K\sin(\psi) \cos^2(\psi) \\
    \times \Big[1 + \frac{L_{p, orb}}{L_{*, spin}}\cos(\psi)\Big],
\end{split}
\end{equation}
where $L_{*, spin}$ is the stellar spin angular momentum, $\Omega_*$ is the star's spin frequency, $k_{2}$ is the planet's Love number, $Q_{10}$ is the tidal quality factor for the obliquity tide, and $\Omega_K$ is the Keplerian orbital angular frequency \citep{lai2012tidal, spalding2022tidal}. We use $L_{*, spin}=L_{CZ}$ to consider the most conservative case in which only the convective zone realigns with the companion orbit. If the convective zone is not decoupled from the stellar core, then the timescale for realignment would be longer due to the larger value of $L_{*, spin}$.

We numerically integrate this expression from $t=30$ Myr (a rough starting age for a K-type star to enter the main sequence) to $t=1$ Gyr (the age of the system) to determine the maximum initial $\psi$ such that the system would today be observed to be aligned at $\psi=1\degree$. The ratio $k_{2}/Q_{10}$ remains poorly constrained and provides the limiting uncertainty within our model. 

\begin{figure}
    \centering
    \includegraphics[width=0.46\textwidth]{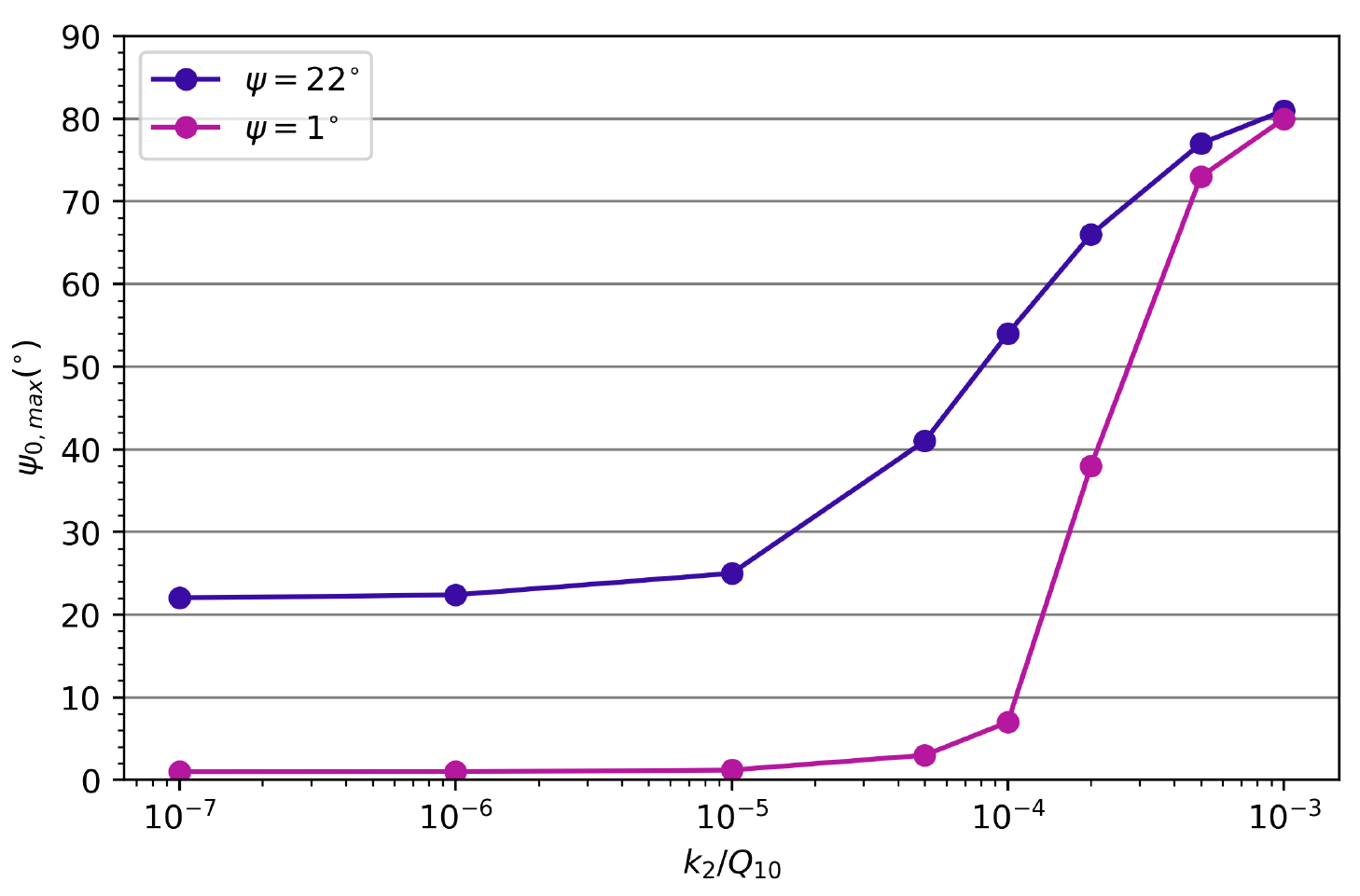}
    \caption{Maximum starting stellar obliquity at the time of protoplanetary disk dispersal, for a range of $k_{2}/Q_{10}$ values under the framework of dynamical tides. We consider the cases in which the true obliquity $\psi$ is either within $1\degree$ of alignment ($\psi=1\degree$) or is set to $\psi=22\degree$ -- the central value determined in Section \ref{section:spinorbitmodel}.}
    \label{fig:dynamical_tides_constraint}
\end{figure}

We consider a range of $k_{2}/Q_{10}$ values in Figure \ref{fig:dynamical_tides_constraint}, where we include both the case in which Qatar-6 A b has a true obliquity $\psi=22\degree$ and the case in which the true obliquity is within $1\degree$ of alignment ($\psi=1\degree$). As demonstrated in Figure \ref{fig:dynamical_tides_constraint}, a relatively large value of $k_{2}/Q_{10}\gtrsim 10^{-4}$ would be required to push Qatar-6 A b from a highly misaligned state to its currently observed spin-orbit angle over the system lifetime. 

Empirical measurements of tidal dissipation in hot Jupiter hosts typically range from $k_2/Q=10^{-5}$ to $10^{-7}$ \citep{penev2018empirical}, with larger values for wider-orbiting planets. The solar system's much wider-orbiting Jupiter, for comparison, has been measured with $k_2/Q=(1.102\pm0.203)\times10^{-5}$ \citep{lainey2009strong}. While the effective tidal quality factor $Q$ differs for each mode of tidal dissipation, $Q$ values for hot and warm Jupiters are typically expected to be high, leading to correspondingly low $k_2/Q$ values. As a result, unless tidal dissipation in warm Jupiter systems is more efficient than previous estimates have suggested, it is unlikely that Qatar-6 A b has been realigned from a large previous misalignment over the post-disk system lifetime.

\subsection{Tidal Circularization Timescale}
\label{subsection:tidal_circ_timescale}
Our model demonstrates that Qatar-6 A b may lie on a slightly eccentric orbit, with $e_b=0.051^{+0.032}_{-0.030}$ (Section \ref{section:spinorbitmodel}). Over time, the orbit will evolve along a path of constant orbital angular momentum characterized by 
\begin{equation}
    a_{b, \mathrm{final}} = a_b (1 - e_b^2),
\end{equation} 
towards $e\rightarrow0$ as energy is removed from the system through tidal dissipation within the planet. As a result, the orbit will ultimately settle to a slightly smaller separation if it currently has a true nonzero eccentricity.

We follow the formulation of \citealt{rice2022origins} to calculate the timescale $\tau_{\rm circ}$ for orbital circularization, with methods summarized here for convenience. $\tau_{\rm circ}$ can be characterized as

\begin{equation}
\tau_{\rm circ}\sim e_b/(de_b/dt),
\end{equation}
 
where

\begin{equation} 
\frac{de_b}{dt} = \frac{dE}{dt}\frac{a_b(1-e_b^2)}{GM_* M_b e_b}\,
\label{eq:dE_dt}
\end{equation} 

and 

\begin{equation}
\frac{dE}{dt} = \frac{21k_2 GM_*^2 \Omega R_b^5}{2Qa_b^6}\zeta(e_b)
\end{equation}
for an incompressible, synchronously rotating planet. Here, $G$ is the gravitational constant, $E$ is the orbital energy of the planet, $Q$ is the planet's effective tidal dissipation parameter, $k_2$ is the planet's Love number, and $\Omega$ is the pseudosynchronous rotation rate, given by

\begin{equation}
    \Omega = \frac{1 + \frac{15}{2}e_b^2 + \frac{45}{8}e_b^4 + \frac{5}{16}e_b^6}{(1 + 3e_b^2 + \frac{3}{8}e_b^4)(1 - e_b^2)^{3/2}}n_b
\end{equation}
for planetary mean motion $n_b$. The corrective factor $\zeta(e)$ in Equation \ref{eq:dE_dt} was derived in \citealt{wisdom2008tidal} and is defined as


\begin{equation} 
\zeta(e_b) = \frac{2}{7}\Big[\frac{f_0(e_b)}{\beta^{15}} - \frac{2f_1(e_b)}{\beta^{12}} + \frac{f_2(e_b)}{\beta^9}\Big],
\end{equation}
where

\begin{equation} 
f_0(e_b) = 1 + \frac{31}{2}e_b^2 + \frac{255}{8}e_b^4 + \frac{185}{16}e^6 + \frac{25}{64}e_b^8
\end{equation} 

\begin{equation} 
f_1(e_b) = 1 + \frac{15}{2}e_b^2 + \frac{45}{8}e_b^4 + \frac{5}{16}e_b^6
\end{equation} 

\begin{equation}
f_2(e_b) = 1 + 3e_b^2 + \frac{3}{8}e_b^4
\end{equation} 

\begin{equation} 
\beta = \sqrt{1-e_b^2}.
\end{equation}


Considering typical expected ranges of $k_2/Q\sim10^{-5}$ to $10^{-7}$ for close-in giant planets \citep{penev2018empirical}, we obtain $\tau_{\rm circ}\sim10^{7}-10^{9}$ years. Based on these relatively short timescales, we cannot exclude the possibility that Qatar-6 A b, with an age $\tau= (1.0\pm0.5) \times 10^9$ yr \citep{alsubai2018qatar}, began its dynamical evolution at a higher eccentricity that has been damped over time. However, because the planet's orbit is consistent with $e_b=0$ within $2\sigma$, we find no strong evidence requiring that the planet's orbit must have had a higher eccentricity in the past. 


\subsection{Kozai-Lidov Timescale}
\label{subsection:kozai_timescale}

While, in the line-of-sight direction, the orbit of Qatar-6 A b appears to be aligned with the Qatar-6 AB binary system orbit ($i_b\sim i_{B}\sim90\degree$), we cannot rule out the possibility that its orbit may be misaligned in the sky-plane and therefore inclined relative to the Qatar-6 AB binary system. If this is the case, then Qatar-6 A b could be located along a low-eccentricity trough of a Kozai-Lidov cycle, where the z-component of the angular momentum vector

\begin{equation} 
    L_z = \sqrt{1-e_b^2}\cos i_{tot}
\end{equation}
is conserved at the quadrupole level. Here, $i_{tot}$ refers to the true inclination between the Qatar-6 A b warm Jupiter orbit and the Qatar-6 AB stellar binary orbit.

At the quadrupole level of approximation, the Kozai-Lidov timescale in a hierarchical 3-body system is given by

\begin{equation}
    \tau_{KL} = \frac{16}{30\pi}\frac{P_2^2}{P_1}(1 - e_2^2)^{3/2} \Big(\frac{m_1 + m_2 + m_3}{m_3}\Big),
\end{equation}
where subscripts 1 and 2 refer to the inner and outer orbits, respectively \citep{naoz2016eccentric}. In our case, $m_1 << m_2, m_3$ such that we can approximate $(m_1 + m_2 + m_3) \sim (m_2 + m_3)$. Simplifying our general expression and reconfiguring it for our system, we obtain

\begin{equation}
    \tau_{KL} \approx \frac{16}{30\pi}\frac{P_{B} ^2}{P_b}(1 - e_{B}^2)^{3/2} \Big(\frac{M_{A} + M_{B}}{M_{B}}\Big).
\end{equation}
The primary star Qatar-6 A is referred to here with the subscript $A$, whereas the orbital properties of the stellar binary companion Qatar-6 B are denoted by the subscript $B$.

From our LOFTI orbital fitting results, we adopt a semimajor axis of $a_{B}=5.36\arcsec$, which translates to $a_{B}=541$ au at a distance of 100.95 pc measured through the parallax reported in \textit{Gaia} DR3. Given our poor constraint on the orbital eccentricity of the stellar binary system, we approximate this timescale for a range of possible eccentricities from $e=0$ to $e=0.9$. We find a timescale ranging from $\tau_{KL}\sim10^{9}$ yr for $e=0.9$ to $\tau_{KL}\sim10^{10}$ yr for $e=0$. These timescales are comparable to the estimated age of the system ($(1.0\pm0.5) \times 10^9$ yr \citep{alsubai2018qatar}), indicating that Kozai-Lidov migration likely has not played a major role in the evolution of this system if Qatar-6 A b formed near its current location. The same planet would have a much shorter Kozai-Lidov timescale if it instead formed on a wider orbit (ranging from $\tau_{KL}\sim10^{7}$ yr for $e=0.9$ to $\tau_{KL}\sim10^{8}$ yr for $e=0$ if the planet began with an orbital period $P_b=1$ year), such that migration through Kozai-Lidov orbital evolution may have occurred in the past.

\subsection{Apsidal Precession Timescale}
\label{subsection:apsidal_timescale}
Kozai-Lidov oscillations can be suppressed by additional perturbations that produce apsidal precession at a rate more rapid than that of the Kozai-Lidov mechanism, reducing the orbit-averaged torque induced by the companion star \citep{holman1997chaotic, wu2003planet}. We consider, in particular, the timescale $\tau_{GR}$ for apsidal precession due to general relativity 

\begin{equation}
    \tau_{GR} = \frac{2\pi c^2 (1 - e_b^2) a_b^{5/2}}{3 (GM_{*})^{3/2}},
\end{equation}
which is relatively short for short-period giant planets. In this expression, $c$ is the speed of light. 

For Qatar-6 A b, the timescale for precession induced by general relativistic effects is only $\tau_{GR}=2\times10^4$ years -- much shorter than the system's current Kozai-Lidov timescale $\tau_{KL}\sim10^9 - 10^{10}$ yr (Section \ref{subsection:kozai_timescale}). We can, therefore, rule out the possibility that the Qatar-6 system is currently undergoing Kozai-Lidov oscillations. 

A past mutual inclination of $i_{\rm tot}\geq39.2\degree$ would have been required between the planetary orbit and the binary star orbit to initiate Kozai-Lidov oscillations. In the case of the Qatar-6 system, a significant mutual inclination could remain undetected within the sky-plane direction if Kozai-Lidov oscillations occurred in the past. Alternatively, the system may have settled at a low final mutual inclination after undergoing Kozai-Lidov cycles (as in, e.g., some of the systems simulated by \citealt{naoz2012formation}).

\subsection{Precession Timescales}
\label{subsection:orb_elem_timescale}
The stellar binary companion, Qatar-6 B, provides a small disturbing force

\begin{equation}
    dF = \bar{R}\hat{r}+\bar{T}\hat{\theta}+\bar{N}\hat{z}
\end{equation}
that perturbs the orbit of Qatar-6 A b, altering the observed orbital elements of the system. Here, $\bar{R}$, $\bar{T}$, and $\bar{N}$ are the radial, tangential, and normal components of the force induced by the perturber. In particular, a nonzero mutual inclination between the planet's and the stellar binary's orbits should induce nodal and inclination precession in the orbit of Qatar-6 A b that may manifest as observable transit duration variations (TDVs) in the system. Our best-fitting solution includes a $\sim$5$\degree$ mutual inclination between the planetary orbit ($i_b=85.95\pm0.24\degree$) and the stellar binary orbit ($i_B=90.17^{+1.07}_{-1.06}\degree$).

The rates of nodal ($d\Omega_b/dt$) and inclination ($di_b/dt$) precession under the influence of a perturbing force $dF$ are given by \citep{murray1999solar}



    

\begin{equation}
    \frac{di_b}{dt} = \sqrt{\frac{a_b(1-e_b^2)}{G(M_A+M_b)}}\frac{\bar{N}\cos(\omega_b + f_b)}{1 + e_b\cos f_b}
    \label{eq:Idot}
\end{equation}
and

\begin{equation}
    \frac{d\Omega_b}{dt} = \sqrt{\frac{a_b(1-e_b^2)}{G(M_A+M_b)}}\frac{\bar{N}\sin(\omega_b+f_b)}{\sin i_b (1 + e_b\cos f_b)},
    \label{eq:bigomegadot}
\end{equation}
where $f_b$ is the true anomaly of the planetary orbit, and both precession timescales are driven by the normal component of the perturbing force. We consider the range $f_b\in (0, 2\pi)$ and find the largest possible nodal and inclination precession rates across this range, adopting a mutual inclination $5\degree$ between the planetary orbital plane and the stellar binary orbital plane and using the values derived in this work (Table \ref{table:results}). We obtain precession rates $d\Omega_b/dt=9.6\times10^{-4}$ deg/yr and $di_b/dt=8.5\times10^{-5}$ deg/yr, corresponding to a projected transit duration change of $<1$ minute over the course of a decade. Thus, we do not expect to observe significant transit duration variations caused by the binary perturber Qatar-6 B.





\subsection{Implications of Dynamical Timescales}
\label{subsection:timescale_implications}
Together, the tidal alignment, tidal circularization, Kozai-Lidov, and apsidal precession timescales jointly indicate that the warm Jupiter Qatar-6 A b likely formed quiescently. The system's long tidal alignment timescale suggests that Qatar-6 A b formed within a protoplanetary disk that was primordially aligned with the Qatar-6 A host star, and the system was not dramatically altered from that point to push the system out of alignment. 

The planet's short tidal circularization timescale prevents us from ruling out an initially eccentric orbit within the plane of that protoplanetary disk. The planet's current low-eccentricity orbit is consistent with either an initially circular orbit or a previously higher-eccentricity orbit that was not pushed to a high inclination relative to the stellar spin axis. A higher eccentricity could have been previously excited within the plane of the protoplanetary disk through planet-disk interactions \citep{goldreich2003eccentricity}, planet-planet scattering \citep{rasio1996dynamical, chatterjee2008dynamical}, or resonance crossings with a planetary companion \citep{chiang2003}.

Lastly, the Kozai-Lidov timescale for the Qatar-6 system is comparable to or longer than the age of the system, depending on the true Qatar-6 AB stellar binary orbital properties. If the planet formed near its currently observed orbit, this indicates that the Kozai-Lidov mechanism has likely not played a major role in the evolutionary past of the Qatar-6 AB system. The possibility that the planet is currently undergoing Kozai-Lidov oscillations can also be ruled out based on the short timescale for general relativistic-induced precession within the system. If Qatar-6 A b instead formed on a wider orbit and migrated inward over time, it is possible that the system may have experienced Kozai-Lidov oscillations in the past. To trigger Kozai-Lidov oscillations, however, the system would have needed a past mutual inclination of $i_{\rm tot}\geq39.2\degree$ between the planetary orbit and the binary star orbit.

The line-of-sight orbit-orbit alignment demonstrated in this work, together with the observed evidence for spin-orbit alignment, suggests a fully quiescent formation mechanism: if the system formed with no large mutual inclinations, the requirement of $i_{\rm tot}\geq39.2\degree$ to initialize Kozai-Lidov oscillations would have never been met, even in the case that Qatar-6 A b began as a much wider-orbiting planet that migrated inwards over time. We conclude that Qatar-6 A b most likely reached its current orbit quiescently, either \textit{in situ} or through disk migration.

\section{Potential Causes of Orbit-Orbit Alignment}
\label{section:formation}

\subsection{Binary Formation Scenarios}

There are a few potential avenues through which a binary star system could form, some of which are more or less likely to produce the observed line-of-sight orbit-orbit alignment. Three key mechanisms for binary star formation are dynamical capture, disk fragmentation, and turbulent fragmentation. In this section, we examine the likelihood of each of these scenarios in the context of the Qatar-6 AB binary system.

In the dynamical capture scenario, the binary companion would have been captured by the gravitational potential of the primary star and its protoplanetary disk \citep[][]{Tokovinin2017WideBinaryCapture}. This formation mechanism produces a relatively high rate of highly misaligned systems, with the orientation of the final system set by the impact parameter of nearby passing stars. Consequently, it is unlikely that the observed line-of-sight orbit-orbit alignment would arise directly from dynamical capture. Furthermore, dynamical capture tends to produce much wider binaries ($a>10^4$ au) and requires a series of specific conditions that must be satisfied, including (1) the presence of a third companion or a highly dissipative disk/envelope system to remove kinetic energy from the binary such that the binding energy of the system becomes negative; (2) a nearby star that falls into a specific range of appropriate relative velocities; and (3) an impact angle that is conducive to capture. Because of these conditions, dynamical capture is thought to be relatively rare in systems with a sub-solar-mass primary star \citep{clarke1991star, heller1995encounters, moeckel2007capture}. Therefore, it is plausible but unlikely that the Qatar-6 AB system was produced by the capture of a companion star.

Alternatively, Qatar-6 AB may have formed through disk fragmentation, which can naturally produce binary star systems with primordially aligned protoplanetary disks. In this framework, a massive, gravitationally unstable circumprimary disk produces a stellar companion that inherits its angular momentum vector \citep[][]{Addams1989diskinst, BonnellBate1994binaryform}. However, disk fragmentation is expected to produce relatively close-in companions \citep[$a \lesssim 200$ au; e.g.,][]{Krumholz2007,Tobin2016} with separation bounded by the extent of the circumprimary disk. A distant stellar companion with sky-projected separation $s=482$ au would, accordingly, be unexpected for the relatively low-mass ($M_A=0.829M_{\odot}$) Qatar-6 A host star within the disk fragmentation framework.

A third possibility is that the binary system formed through turbulent fragmentation \citep[][]{Offner2010, Offner2016, Lee2017}, where a gravitationally unstable over-density already in the contraction phase further fragments into two separate stars. While the turbulent environment can imprint potentially misaligned angular momentum vectors onto the stars, the overall angular momentum of the contracting over-density is expected to preferentially produce relatively aligned systems with $i_{tot}\lesssim 45\degree$ \citep[][]{Bate2018}. In this case, the Qatar-6 A b planet could have quiescently formed within the relatively aligned protoplanetary disk, producing either a primordial orbit-orbit alignment or a relatively small primordial misalignment.


\subsection{Dynamical Orbit-Orbit Alignment During Binary-Driven Disk Precession}

If a primordial misalignment existed between the circumprimary disk and the companion star's orbit, the binary potential would drive disk precession about the binary orbit normal \citep[][]{Bate2000, batygin2012primordial, Zanazzi2018}. The gas-rich disk can dissipate the energy available from precession into heat, and in the process the disk can be pushed toward alignment with the binary orbit \citep[][]{papaloizou1995dynamics, Bate2000, lubow2000tilting}. As the associated timescale of this mechanism is well within the typical lifetime of protoplanetary disks for binary separations of order $500$ au \citep[][]{christian2022possible}, dynamical alignment via energy dissipation during binary-driven disk precession could robustly explain the sky-projected inclination match between the binary orbit and transiting warm Jupiter orbit in the Qatar-6 system. 


On the other hand, binary-driven disk precession has been invoked numerous times to explain observed spin-orbit misalignments \citep[][]{batygin2012primordial}. Indeed, the alignment scenario described above would, naively, leave the angular momentum vector of the star unchanged, and thus, given the long tidal realignment timescale in Equation \eqref{eq:tau}, produce a spin-orbit misalignment of order the initial binary-disk misalignment. Our constraint on the true spin-orbit angle $\psi=21.82^{+8.86}_{-18.36}\degree$ leaves room for a non-negligible spin-orbit misalignment that could have resulted from a dynamical forcing of an initially misaligned protoplanetary disk into alignment with the binary orbit. 

However, a true joint spin-orbit and orbit-orbit alignment could naturally arise from dynamical orbit-orbit alignment paired with additional gravitational and magnetic processes that occur over the disk lifetime. During the embedded phase of star formation, accretion and gravitational star-disk coupling can efficiently transfer angular momentum between the star and the disk, suppressing the production of large spin-orbit misalignments \citep{spalding2014alignment}. Furthermore, strong magnetic torques in young systems with relatively low-mass stars ($M_*\lesssim1.2M_{\odot}$) can push systems with primordial spin-orbit misalignments back to alignment within the protoplanetary disk's lifetime \citep{spalding2015magnetic}. As a result, dynamical orbit-orbit alignment during the protoplanetary disk phase does not necessarily preclude spin-orbit alignment.

\section{Conclusions}
\label{section:conclusions} 

In this work, we have demonstrated that all current lines of evidence are consistent with a quiescent formation mechanism for the Qatar-6 system, which includes two stars and one transiting warm Jupiter. Our results are summarized by two key points:

\begin{itemize}
    \item The warm Jupiter Qatar-6 A b is consistent with spin-orbit alignment along the host star's equator, with a project spin-orbit angle $\lambda=0.1\pm2.6\degree$ and a true spin-orbit angle $\psi=21.82^{+8.86}_{-18.36}\degree$. 
    \item Both the planet's orbit ($i_b=85.95\pm0.24\degree$) and the stellar binary orbit ($i_{B}=90.17^{+1.07}_{-1.06}\degree$) are edge-on, such that all three bodies are consistent with alignment in the line-of-sight direction. 
\end{itemize}

We have precisely measured the spin-orbit alignment of Qatar-6 A b within the 2D sky plane, and we find that the planet's 3D spin-orbit angle is also consistent with alignment (albeit with larger uncertainties). Interestingly, our results further suggest a joint orbit-orbit alignment across the three-body system: both the transiting planet and the stellar binary lie in an edge-on configuration. Such a 3D alignment may have been produced either primordially or through dynamical alignment of the protoplanetary disk and quiescent formation of the warm Jupiter within that disk. The full 3D alignment of the system cannot be confirmed due to the unconstrained transit direction within the sky plane. 


This system offers a detailed case study with multiple lines of evidence pointing toward a likely quiescent formation mechanism. The gaps in data pervasive in studies of individual systems, such as the one presented here, may be possible to remediate by applying statistical arguments to a wider sample of planetary systems. Future studies examining population-wide trends in the spin-orbit and orbit-orbit orientations of binary systems will offer further insights into the key dynamical mechanisms that dominate the evolution of exoplanet systems.

\section{Acknowledgments}
\label{section:acknowledgments}

We thank the anonymous referee for their helpful comments that have improved this manuscript. We also thank Andrew Vanderburg and Sam Christian for helpful discussions, and Sam Yee for providing support for our Keck/HIRES observations. M.R. thanks the Heising-Simons Foundation for their generous support. This work is supported by the Astronomical Big Data Joint Research Center, co-founded by National Astronomical Observatories, Chinese Academy of Sciences and Alibaba Cloud. 

The data presented herein were obtained at the W. M. Keck Observatory, which is operated as a scientific partnership among the California Institute of Technology, the University of California, and the National Aeronautics and Space Administration. The Observatory was made possible by the generous financial support of the W. M. Keck Foundation. The authors wish to recognize and acknowledge the very significant cultural role and reverence that the summit of Maunakea has always had within the indigenous Hawaiian community.  We are most fortunate to have the opportunity to conduct observations from this mountain.  

This research has made use of the Keck Observatory Archive (KOA), which is operated by the W. M. Keck Observatory and the NASA Exoplanet Science Institute (NExScI), under contract with the National Aeronautics and Space Administration. This research has also made use of the NASA Exoplanet Archive, which is operated by the California Institute of Technology, under contract with the National Aeronautics and Space Administration under the Exoplanet Exploration Program.

\software{\texttt{allesfitter} \citep{gunther2021allesfitter}, \texttt{emcee} \citep{foremanmackey2013}, \texttt{isoclassify} \citep{huber2017isoclassify, huber2017asteroseismology, berger2020gaia}, \texttt{lightkurve} \citep{cardoso2018lightkurve}, \texttt{lofti\_gaia} \citep{pearce2020orbital}, \texttt{matplotlib} \citep{hunter2007matplotlib}, \texttt{numpy} \citep{oliphant2006guide, walt2011numpy, harris2020array}, \texttt{pandas} \citep{mckinney2010data}, \texttt{scipy} \citep{virtanen2020scipy}, \textit{The Cannon} \citep{ness2015cannon}}

\facility{Keck: I (HIRES), Exoplanet Archive, Extrasolar Planets Encyclopaedia}

\bibliography{bibliography}
\bibliographystyle{aasjournal}

\end{document}